# Towards nanomechanical models of liquid-phase exfoliation of layered 2D nanomaterials: analysis of a $\pi - peel$ model

Lorenzo Botto[1*,2]

[1]Process & Energy Department, TU Delft, 2628 CB, Delft, The Netherlands
[2]School of Engineering and Materials Science
Queen Mary University of London, E1 4NS, London, UK
[*]Email: l.botto@tudelft.nl

Abstract: In liquid-phase exfoliation for the production of 2D nanomaterials fluid forces are used to gently overcome adhesive interlayer forces, leading to single- or few-layer 2D nanomaterials. Predicting accurately the critical fluid shear rate for exfoliation is a crucial challenge. By combining notions of fluid mechanics and fracture mechanics, we analyse a mathematical model of exfoliation, focusing on the $\pi - peel$ regime in which bending forces are much smaller than the applied hydrodynamic forces. We find that in this regime the shear rate is approximately proportional to the adhesion energy, independent of the bending rigidity of the exfoliated sheet, and inversely proportional to the size $a$ of a (assumed pre-existing) material flaw. The model appears to give values comparable to those obtained in wet ball milling, but to overestimate the shear rate values reported for turbulent exfoliation (by rotor mixing or microfluidization). We suggest that for turbulent exfoliation a "cleavage model" may be more appropriate, as it gives a stronger dependence on $a$ and smaller critical shear rates.

1. Introduction

Graphene and other 2D nanomaterials in the form of atomically thin nanosheets promise unexpected performance in many applications. The nanosheets can be embedded in nanocomposites to make them conductive (Santagiuliana, 2018), improve barrier properties or increasing strength and toughness (Rafiee, 2010 ). Or they can be suspended in liquid solvents to produce conductive inks for printed electronics and high-performance coatings (Torrisi, 2018). While these applications are currently tested in small scale applications, to reach true market impact it is paramount to produce large quantities of 2D nanosheets cheaply, and with control over thickness, lateral area and amount of defects (Ferrari, 2015).

A very promising technique for the large-scale production of 2D nanosheets is liquid-phase exfoliation (Coleman, 2009; Yi, 2015). This technique is relatively simple. It consists in subjecting microparticles of layered 2D nanomaterial (each microparticle being composed of hundreds or thousands of layers) to large mechanical forces that detach the layers. Several different variants of the technique exist, the main ones being turbulent exfoliation (e.g. in rotor-stator mixers (Paton, 2014) or microfluidisation devices (Paton, 2017)), wet ball milling (Knieke, 2010) (Zhao, 2010), and sonication (Alaferdov, 2014). These techniques have in common the fact that the microparticles are initially suspended in a liquid (in wet ball milling the particles are wet by a liquid, but a thin layer of liquid is still present). In addition to transmit mechanical stresses, the liquid enables to reduce the adhesion between the layers, prevent reaggregation, and make the mechanical action less aggressive (Shen, 2015). Bottom-up synthesis methods, such as Chemical Vapour Deposition in its different variants (Aïssa, 2015), are promising for producing high-quality 2D nanomaterials particularly suitable for devices. However, to produce 2D nanomaterials very cheaply for large-scale applications such as nanocomposites, inks or coatings, liquid-phase exfoliation is difficult to beat.

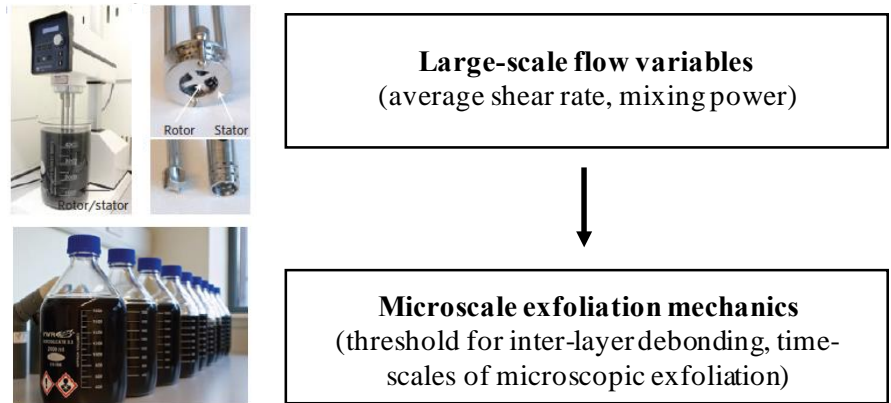

*Figure 1: Optimising liquid-phase exfoliation processes (left, reproduced with permission from Ref. (Paton, 2014)) requires models to link large-scale flow variables to the micromechanics of exfoliation.*

Optimising liquid-exfoliation processes requires addressing a delicate balance: the mechanical stresses applied to the particles by the fluid must be sufficiently high to delaminate the particles, but not much higher. Excessive stresses can fragment the nanosheets, producing small area sheets of low intrinsic value, or damage the sheets (Johnson, 2015). Reaching the right stress level is thus paramount. However, particle-level stresses cannot be controlled directly. What can be controlled are large scale flow variables, such as the mixing power or, equivalently, the average shear rate the suspension is subject to. These are the macroscopic variables that can be controlled by the user in the production process. It would be highly valuable if analytical formulas relating these macroscopic variables to microscopic exfoliation thresholds and time-scales were available. Developing these formulas requires an understanding of the flow physics and deformation mechanics at the particle level (Fig. 1).

Quite surprisingly, despite the growing importance of liquid-phase processing in the production of graphene and other 2D nanomaterials, the development of theoretical models for liquid-phase exfoliation is at its infancy. Two theoretical models have appeared recently which seem to be relevant. An exfoliation model based on a sliding deformation was proposed by Chen et al. (Chen, 2012), and later extended by Paton et al (Paton, 2014). In this model, the shear forces exerted by the fluid are assumed to balance the rate at which the total surface energy (including solid-solid, solid-liquid and liquid-liquid interaction energy) changes with respect to the sliding distance. The model was used to describe the dependence of the critical shear rate for exfoliation on the size of the suspended plate-like particles, the viscosity of the fluid, and the energy of adhesion. Chen-Paton's model is insensitive to the mechanical property of the particle: the bending rigidity or young Modulus of the particles do not appear in the model. The assumptions of this model were not stated with sufficient clarity to rigorously assess its validity from a nanomechanical perspective. Prior to the work of Chen, in 2009, a model of exfoliation was developed by Borse et al. (Borse, 2009) to model the exfoliation of multilayer clay particles in polymers. Unlike Chen/Paton's model, Borse's model is sensitive to the mechanical properties of the particle. The formulation of Borse's model is based on Kendall's theory for peeling of elastomers (Kendall, 1975). Borse correctly identified that because the area of contact between the sheets is large, a simple balance between adhesive forces and shear forces gives values of the shear stress too high to be realistic. As a consequence, one must hypothesise that the debonding of the layers is due to stress concentration at the microscopic crack tip of a pre-existing flaw in the inter-layer interface.

Borse's model includes in the fracture mechanics formulation the work done by the fluid forces, the stretching energy associated to the extension of each layer, and the adhesion energy associated to the van der Waals forces between the layers. For the case of constant edge load

(identical to the one analysed by Kendall), Borse and collaborators analysed the values of the critical shear stress as a function of the peeling angle and the Young modulus of the exfoliated sheet.

We adopt a view similar to that of Borse, and consider exfoliation due to extension of an initial flaw in the inter-layer interface. The extension is caused by a peeling process, whereby the forces driving peeling are hydrodynamic in nature. We envision that the fluid opens a pre-existing crack of length $a$. The opening angle will depend on the ratio of hydrodynamic and bending forces. If this ratio is small, the opening angle will be small. If this ratio is much larger than one, the flap will turn by 180°, and the direction of pulling will be parallel to the direction of propagation of the crack, as illustrated in Fig. 2b. This configuration is analogous to the one considered in the "$\pi - peel'$ mechanical test to measure adhesion (Lin, 2002). For brevity, in the current paper, we refer to this configuration as "$\pi$-peel" configuration.

In this paper we analyse in detail a mathematical model of exfoliation for this "$\pi$-peel" configuration, by rigorously justifying the fluid and solid mechanics aspects of the problem. Particularly, we discuss the parameter values for which this configuration may be observed. In contrast to the model by Borse et al., we assume that each layer is perfectly inextensible. Our work takes inspiration from the work on the "inextensible fabric" approximation discussed in Ref. (Roman, 2013), but we recast our results in the context of 2D nanomaterials processing, so that a direct comparison with experimental data from the literature can be made.

The problem we are tackling is one of the first explorations in terms of mechanics of a very complex fluid-structure interaction problem, which needs to be analised from different perspectives to be fully understood. Rather than analysing in depth a sophisticated model, our aim with the paper is to test the prediction of simple models which will enable us to identify the theoretical directions that could bring us close, in terms of orders of magnitude, to the published experimental data for the critical shear rates.

2. **Dimensional analysis of the exfoliation problem**

Before considering a particular exfoliation configuration, let us first analyse the general problem of exfoliation from the point of view of dimensional analysis. The critical shear rate for exfoliation, $\dot{\gamma}$, depends on the properties of the fluid (density $\rho$, viscosity $\mu$, and wetting properties), the mechanical and geometrical properties of the layered micro-particle (particle length $L$, particle width $W$, Young modulus of each layer $E$, bending rigidity of each layer $D_0$, total number of layers $N$, flaw size $a$), and the mechanical and geometrical properties of each inter-layer interface (adhesion energy $\Gamma$, area of contact, presence of flaws, etc.).

In exfoliation problems, the particle Reynolds number is typically small, $Re = \rho\dot{\gamma}L^2/\mu \ll 1$, so $\rho$ is not an important parameter. Among the mechanical properties, the adhesive properties of each inter-layer, parameterised by $\Gamma$ and the bending rigidity of each layer $D_0$ are likely to be dominant controlling parameters. The Young modulus of graphene is huge ($E \sim 1TPa$ (Lee, 2008)) while its bending rigidity is low ($D_0 \sim 7eV$ (Lindahl, 2012)), so we may regard graphene as an inextensible membrane of finite bending rigidity.

The above parameters suggest a functional relationship of the form

$$\dot{\gamma} = f(\Gamma, D_0, N, n, \mu, a, L, W) \quad (1)$$

Some simplifications are possible, under the following hypotheses and observations:

- In the zero Reynolds number limit appropriate for colloidal particles, the pressure and viscous stresses scale as $\mu\dot\gamma$, so the viscosity enters into the problem only multiplied by the shear rate;
- It can be assumed that the forces induced by the fluid scale proportionally to $W$. This is not strictly true unless $W \gg L$, while typically $W \sim L$. However, if edge effects are neglected this is a reasonable approximation if the goal is to obtain order of magnitude estimates;
- The direct dependence on $n$ can be neglected if one assumes $\frac{n}{N} \ll 1$. The indirect dependence of the problem on $n$ is still present, because the bending rigidity of the flap depends on $n$. The bending rigidity of multilayer graphene scales roughly as $n^3$ for $n \geq 3$, so for multilayers we can write $D(n) \cong D_0 n^3$, where $D_0 \cong 20 eV$ is the extrapolated value of the single graphene sheet (Sen, 2010);
- The dependence on $L$ can be ignored, at least as a first approximation, because the applied hydrodynamic forces on the flap and the "resistive" forces due to elasticity and adhesion depend primarily on $a$.

Under the hypotheses above, using $a$ and $D$ to make the other parameters dimensionless Eq. (1) can be written as

$$\frac{\dot\gamma \mu a^3}{D} = f_1\left(\frac{\Gamma a^2}{D}\right) \quad (2)$$

where $f_1$ is a non-dimensional function of its argument. Equation (2) shows that the *non-dimensional critical shear rate*, $\frac{\dot\gamma \mu a^3}{D}$, is a unique function of the *non-dimensional adhesion energy*, $\frac{\Gamma a^2}{D}$.

To discuss the comparison of analytical and experimental results, it is useful to assume for $f_1$ a power-law relationship, for which Eq. (2) becomes

$$\frac{\dot\gamma \mu a^3}{D} \sim \left(\frac{\Gamma a^2}{D}\right)^\xi \quad (3)$$

The values the exponent $\xi$ can attain are constrained by physical consideration. It is expected that an increase in flaw size will make the material weaker, so $\dot\gamma$ must decrease with increasing $a$. This requires $\xi < 3/2$. In addition, an increase in adhesion energy must translate into a larger fluid shear stress. This is only possible if $\xi > 0$.

An interesting possibility is the case $\xi = 1$. For this choice of the exponent, the bending rigidity becomes an irrelevant parameter. A complete independence on the bending rigidity can only be plausible if the bending rigidity is so low, that its precise value does not matter (or nearly so).

## 3. Exfoliation in the "π-peel" configuration

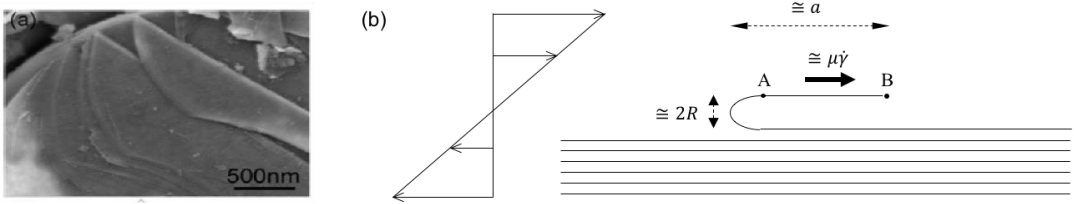

*Figure 2: (a) Peeling deformation of Boron Nitride following wet ball milling; Scanning Electron Microscopy image reproduced with permission from (Li, 2011). b) Schematic of the "π-peel" configuration. We assume that the inter-layer interface is debonded over a length $a - 2R \approx a$, where $R$ is the radius of curvature of the fold. The local shear flow, whose linear profile is illustrated in the sketch, produces a tangential stress on the flap of order $\mu\dot{\gamma}$.*

Figure 2a, from (Li, 2011), shows the surface of a nanomaterials after wet ball milling. Flexible layers of nanomaterials have been partially peeled off due to strong shear forces, leaving a fold of very small radius of curvature. A model for this situation can be developed by considering a continuum sheet partially detached from a rigid "mother particle" (Fig. 2b). We assume that all the sheets have the same length $L$. The total thickness of the microparticle is $h \ll L$. The geometry of the peeled flap is composed of a curved fold, of radius of curvature $R$, and a flat portion (from point A to point B in the schematic). The flat portion is subject to a tangential shear stress $\mu\dot{\gamma}$.

In the case of ball milling, the shear stress is created by a relative velocity $U$ between the milling balls acting on a small gap $d$ between the balls, leading to a shear rate $\dot{\gamma} \cong U/d$. In the case of a multilayer particle suspended in a turbulent flow, the ambient shear rate instantaneously "seen" by the particle is the result of the work done by the largest turbulence structures on the smallest, dissipative Kolmogorov eddies (Tennekes, 1972). For microparticles smaller than the Kolmogorov scale, equilibrium between the energy rate input $P$ (e.g. the mixing power) and the viscous dissipation occurring at the scale of the particles gives $\dot{\gamma} \cong \sqrt{P/(V\mu)}$, where $V$ is the liquid volume (see, e.g., (Varrla, 2014) for an application to graphene).

The length of the detached layer is $a$. Because $R \ll a$, the flap length is also approximately equal to $a$. The tangential force per unit width acting on the flap is $\mu\dot{\gamma}a$.

We need to better specify certain fluid mechanics assumptions. First of all, the model assumes that the particle is aligned with the flow. In fact, plate-like particles rotate when suspended in a shear flow. However, the rotation is very slow when the plate-like particle is nearly aligned with the flow and the aspect ratio is large, with a rate of rotation of the order of $h/L\dot{\gamma}$ (Jeffery, 1922). For most of its rotation period, a plate-like particle of large aspect ratio such as a multilayer microparticle can be considered to be aligned with the flow. A second aspect not included in the model above is the effect of normal hydrodynamic stresses. For a plate-like particle aligned with the flow, normal stresses of the order of $\left(\frac{h}{L}\right)\mu\dot{\gamma}$ act on the surfaces of the particle parallel to the flow direction (e.g. on the flap surface between A and B) (Singh, 2014). Because $\frac{h}{L} \ll 1$, along the surface of the flap these stresses are subdominant with respect to the tangential stresses and are thus neglected in the current model. The normal stresses acting on the curved fold, in the direction of the flow, are instead of $O(\mu\dot{\gamma})$. The effect of normal stresses on the model predictions will be considered in Sec. 3.2. We assume that $L$ is smaller than any scale of the flow, so that the flow field around the particle is smooth. For particles in turbulence, this requirement translates to $L \ll \eta_K$, where $\eta_K$ is the Kolmogorov scale (Landau, 1986).

The critical value of $\dot{\gamma}$ for exfoliation can be calculated by considering the instantaneous equilibrium between the external work, the change in adhesion energy and the change in bending energy for an inextensible sheet. The analysis is similar to that in (Roman, 2013), but now the external force is not constant. The velocity of point A (or B) is twice the velocity of the advancing peeling front. As a consequence, as the peeling front moves by an amount $da$ the work done by the external force is $(W\mu\dot{\gamma}a)2da$, where $W$ is the width. The change in adhesion energy is $W\Gamma da$, where $\Gamma$ is the adhesion energy per unit area (also called work of separation). Calling $dE(\dot{\gamma})$ the infinitesimal change in adhesion energy corresponding to $da$, the critical value of $\dot{\gamma}$ for crack initiation satisfies:

$$dE(\dot{\gamma}) + W\Gamma da = 2W\mu\dot{\gamma}ada \quad (4)$$

The bending energy is denoted $E(\dot{\gamma})$ to highlight that this quantity if a function of the shear rate. Considering a curvilinear coordinate $s$ with origin at the crack tip, the bending energy is $E = \frac{1}{2}DW \int_0^a (\dot{\theta})^2 ds$, where $\theta(s)$ is the rotation angle at $s$ and $\dot{\theta} = \frac{d\theta}{ds}$ is the local curvature. The bending energy integral is dominated by the curvature $1/R$ at the crack tip, which has an extent of the order of $2R$. Thus, the order of magnitude of the bending energy is $W\frac{D}{R}$. The function $\theta(s)$ can be calculated by considering the equation for the Elastica (Audoly, 2010) for a fixed value of $a$, which in our case reads

$$\ddot{\theta} + \frac{\mu\dot{\gamma}a}{D}\sin(\theta) = 0 \quad (5)$$

Here $D$ is the bending rigidity of the elastic element. Equation (5) is obtained from the equation of equilibrium to rotation for an infinitesimal element of flap, $\frac{dM}{ds} = (\boldsymbol{F}(s) \times \boldsymbol{t}) \cdot \boldsymbol{e}_z$ (Landau, 1986). In this expression, $M = D\dot{\theta}$ is the moment of the internal stresses, $\boldsymbol{F}$ is the internal force per unit width, $\boldsymbol{t}$ is the local tangent vector, and $\boldsymbol{e}_z$ is the unit vector oriented along the width direction. We have assumed that the only external (hydrodynamic) forces act from point A to point B. Hence, in the curved portion of the flap, the equation of equilibrium to translation $\frac{d\boldsymbol{F}}{ds} = 0$ requires that $\boldsymbol{F}$ is a constant vector (Landau, 1986). Such constant vector can be easily calculated by noting at point A the force per unit width $\boldsymbol{F}$ is equal to the tension $\cong \mu\dot{\gamma}a\boldsymbol{e}_x$ (the unit vector $\boldsymbol{e}_x$ being parallel to the flow and pointing in the flow direction). As a consequence, $\boldsymbol{F}(s) \cong \mu\dot{\gamma}a\boldsymbol{e}_x$ and $(\boldsymbol{F}(s) \times \boldsymbol{t}) \cdot \boldsymbol{e}_z \cong \mu\dot{\gamma}a \sin\theta$. Inserting this last expression into the equation of equilibrium to rotation gives (5).

The parameter $\frac{\mu\dot{\gamma}a}{D}$ appearing in Eq. (5) has the dimensions of the square of a reciprocal length. Because there are no other characteristic lengths in the problem, we anticipate that the curvature of the fold $R$ scales as $\left(\frac{D}{\mu\dot{\gamma}a}\right)^{1/2}$.

A solvable first-order non-linear equation can be obtained by multiplying Eq. (5) by $\dot{\theta}$, and integrating with respect to $s$. The result is $\frac{1}{2}(\dot{\theta})^2 - \frac{\mu\dot{\gamma}a}{D}\cos(\theta) = c_1$, where $c_1$ is a constant. For values of $s$ corresponding to the region from point A to point B, the rotation angle is constant and equal to $\theta = \pi$. Hence, $c_1 = \frac{\mu\dot{\gamma}a}{D}$. Using this value and the trigonometric identity $(1 + \cos\theta) = 2\cos^2\theta/2$, we obtain $\dot{\theta} = 2\sqrt{\frac{\mu\dot{\gamma}a}{D}}\cos(\frac{\theta}{2})$. This is a separable equation whose solution is $\theta(s') = 2\arcsin(\tanh s')$, where $s' = s/\sqrt{\frac{D}{\mu\dot{\gamma}a}}$. The corresponding bending energy is

$$E = 2W\sqrt{\mu\dot{\gamma}aD} \quad (6)$$

For $s \ll \sqrt{\frac{D}{\mu\dot{\gamma}a}}$ we have $\theta(s) \cong 2s/\sqrt{\frac{D}{\mu\dot{\gamma}a}}$. Hence, the radius of curvature near the crack tip is of the order of $\left(\frac{D}{\mu\dot{\gamma}a}\right)^{1/2}$, as anticipated.

Inserting expression (6) into the energy balance (4) gives

$$a^{-1/2}\sqrt{\mu\dot{\gamma}D} + \Gamma = 2\mu\dot{\gamma}a \quad (7)$$

This expression yields the critical shear rate as a function of the parameters $a$, $D$ and $\mu$.

It is convenient to recast Eq. (7) in terms of the non-dimensional shear rate $\frac{\dot{\gamma}\mu a^3}{D}$ and non-dimensional adhesion energy $\frac{\Gamma a^2}{2D}$, introduced in Sec. 2:

$$\frac{\Gamma a^2}{2D} = \frac{\dot{\gamma}\mu a^3}{D} - \frac{1}{2}\sqrt{\frac{\dot{\gamma}\mu a^3}{D}} \quad (8)$$

In contrast to the constant edge force case (Roman, 2013), in our case the bending energy depends on $a$ and the term $dE$ is not zero, giving rise to the square root term on the right-hand side of Eq. (8).

From Eq. (8), and comparing with Eq. (3), the limit $\frac{\mu\dot{\gamma}a^3}{D} \to \infty$ gives a power-law exponent $\xi = 1$, exactly. The effect of the bending energy term is to increase the critical exfoliation value obtained in this asymptotic limit by an amount that depends on the square-root of the shear rate.

### 3.1 Conditions for "π-peel"

For the analytical solution (8) to be valid, the flap length must be long in comparison to the radius of curvature of the fold (Roman, 2013). The condition $R \ll a$ gives $\frac{\mu\dot{\gamma}a^3}{D} \gg 1$. Eq. (9) also requires the flap geometry at equilibrium to assume a shape similar to that in Fig. 2. For the flap to bend to such an extent the ratio of viscous forces ($\sim \mu\dot{\gamma}Wa$) to bending forces ($\sim DW/a^2$) must be large. This, again, gives $\frac{\mu\dot{\gamma}a^3}{D} \gg 1$. Hence, the condition $\frac{\mu\dot{\gamma}a^3}{D} \gg 1$ simultaneously ensures a good separation of scale between the fold and the flap, and that the assumption of nearly tangential viscous force on the flap is met.

Figure 3 illustrates how the radius of curvature of the fold, estimated as $R \cong \frac{1}{2}\left(\frac{D}{\mu \dot{\gamma} a}\right)^{1/2}$, varies with the length of the flap, for three typical shear rates. Rather than plotting $R$ we plot $2R$, which gives a measure of the maximum height of the folded region. The bending rigidity is set to $D = 10^{-18}$ J, close to that of single-layer graphene (Lindahl, 2012). In Figure 3a, a dynamic viscosity $\mu = 0.001 \, Pa \cdot s$ is assumed, typical of aqueous solvents and NMP (Paton, 2014). In figure 3b, the viscosity is increased to $\mu = 1 \, Pa \cdot s$. For $\mu = 0.001 \, Pa \cdot s$, a good separation of scales between $R$ and $a$ occurs if the shear rate is at least $\dot{\gamma} = 10^6 s^{-1}$. As the viscosity increases to $\mu = 1 \, Pa \cdot s$, the "π-peel" configuration regime occurs for smaller values of shear rate (Fig. 3b).

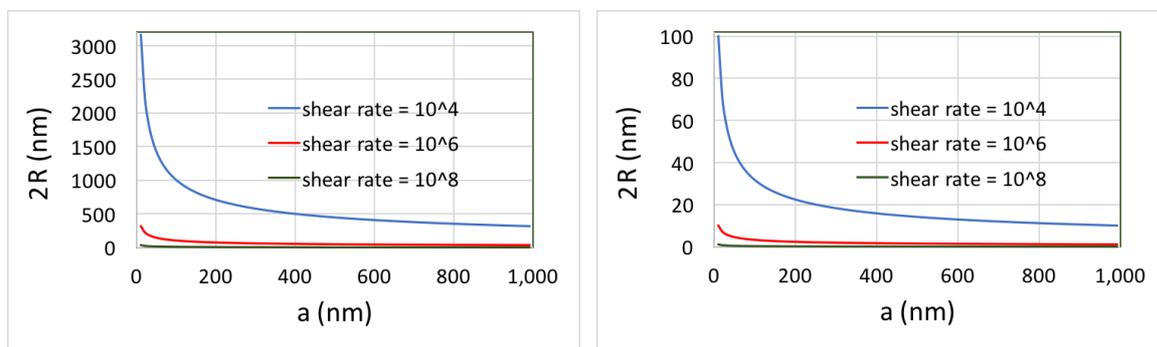

*Figure 3: Two times the radius of curvature of the fold vs. length of the flap for (a) $\mu = 0.001 \, Pa \cdot s$ and (b) $\mu = 1 \, Pa \cdot s$. The bending ridigity is assumed to be $D = 10^{-18} J$, corresponding to approximately 6eV. The quantity 2R corresponds approximately to the maximum height of the fold.*

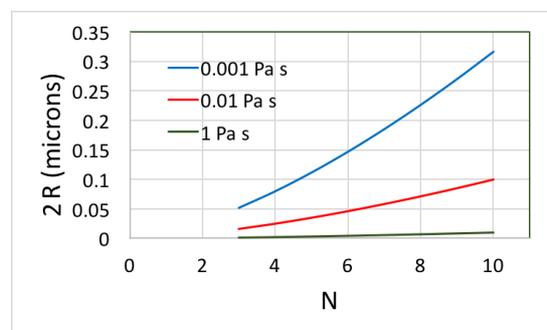

*Figure 4: Two times the radius of curvature of the fold vs. number of layers for a=200nm, $\dot{\gamma} = 10^6 s^{-1}$ and $D = 10^{-18} J$.*

The radius of curvature of the flap increases strongly with the number of layers n. Assuming a cubic relation $D = D_0 n^3$ yields the results shown in Fig. 4. For this figure, a = 200nm and $\dot{\gamma} = 10^6 s^{-1}$; results are shown for three different viscosities. Even considering relatively few layers, for R to be sufficiently small in comparison to a, either the viscosity must be relatively large or, if the viscosity is comparable to that of water, the shear rate must be larger than $10^6 s^{-1}$.

We can call *strongly-stressed* nanosheets, 2D material nanosheets for which the scale separation between $R$ and $a$ is complete, and *mildly-stressed* nanosheets, nanosheets for which $R$ is larger than $a$, but not by a very large factor. Correspondingly we have two approximations for the critical shear rate. For strongly stressed nanosheets the bending rigidity contribution is negligible and

$$\dot{\gamma} \approx \frac{\Gamma}{2\mu a} \quad (9)$$

This equation is, strictly speaking, a good approximation when $\frac{\mu\dot{\gamma}a^3}{D} \to \infty$ (which is equivalent to $\Gamma a^2/D \to \infty$). For mildly stressed sheets, solving Eq. (8) for $\dot{\gamma}$ yields

$$\frac{\dot{\gamma}\mu a^3}{D} = \frac{\Gamma a^2}{2D}\left(1 + \frac{1}{4(\Gamma a^2/D)} + \frac{\sqrt{1 + \frac{8\Gamma a^2}{D}}}{\frac{4\Gamma a^2}{D}}\right) \quad (10)$$

This approximation is more accurate than (9) when $\frac{\Gamma a^2}{D}$ (or equivalently $\frac{\mu\dot{\gamma}a^3}{D}$) is large but finite. For order of magnitude estimates, both (9) and (10) are acceptable. Note that $\Gamma$ is the work required to separate two surfaces, so $\Gamma/2$ is equal to the surface tension of the solid.

Equation (9) is consistent with Kendall's theory for peeling of an extensible thin sheet (Borse's model for exfoliation of clays (Borse, 2009) is based on Kendall's theory, so it reaches the same conclusions). In the case of a constant edge load $F$ applied at an angle $\phi$, Kendall calculated that the peeling force for an elastic film of thickness $d$ and Young's modulus $E$ satisfies $\left(\frac{F}{W}\right)^2 \frac{1}{2dE} + \left(\frac{F}{W}\right)(1 - cos\phi) = \Gamma$. For $E \to \infty$ (inextensible sheet) and $\phi = \pi$, this equation recovers Eq. (9) when $F = \mu a W$. Incidentally, the order of magnitude scaling suggested by Eq. 9 is identical to that obtained by Paton et al. (Paton, 2014) for sliding of parallel, rigid platelets, although their configuration is different. In Paton's formulation, the energy term is given by the change in surface energy corresponding to a change in overlap length $x$ between the platelets. Paton's argument for the obtaining the critical sliding force is that the overlap area is $Wx$, and the corresponding surface energy is $\Gamma W x$ (apart from constant energy terms that do not affect the force). Thus, the sliding force $d/dx(\Gamma x)$ per unit width is constant, and equal to $\Gamma$. This results seems to contradict Kendall's theory, which shows that for $\phi = 0$ (as in a lap shear joint configuration) $F$ is proportional to $\sqrt{\Gamma}$, not $\Gamma$. A possible explanation for the discrepancy is the neglect in Paton's model of the straining of molecular bonds as two rigid sheets slide by a distance comparable to the crystal lattice (Xu, 2011) and the disregard of non-uniformities in the interfacial stress (Pugno, 2010).

*3.2 Normal load on the curved fold*

Models (9) and (10) neglect the normal force on the curved fold acting in the direction of the flow. This distributed force pushes the fold in the same direction as the tangential force pulling the flap, so we expect a reduction in the critical shear rate. We can better quantify this statement. The normal force on the flap $f$ can be estimated to be of the order of $\mu\dot{\gamma}RW$. Without attempting to solve Eq. (5) exactly by including a distributed force, we can approximate the problem by assuming that $f$ is concentrated in the mid-point of the fold, $s \cong R/2$. At this location the internal force $\boldsymbol{F}(s)$ will have a discontinuity: for $s > R/2$, $\boldsymbol{F} \approx \mu\dot{\gamma}a\boldsymbol{e}_x$ as before; for $s < R/2$, $\boldsymbol{F}$ will increase to $\approx \mu\dot{\gamma}(a + R)\boldsymbol{e}_x$.

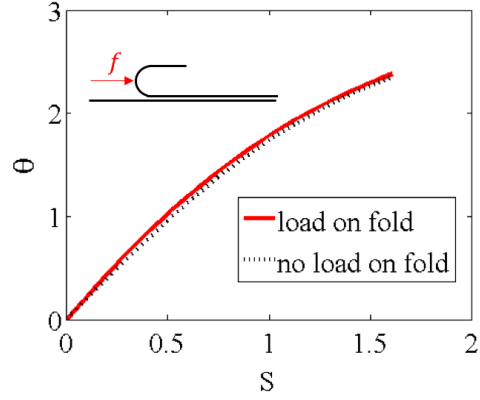

Figure 5: Effect of a point load of magnitude $f = \mu R \dot{\gamma} W$ acting on a fold of radius R for R/a=0.7

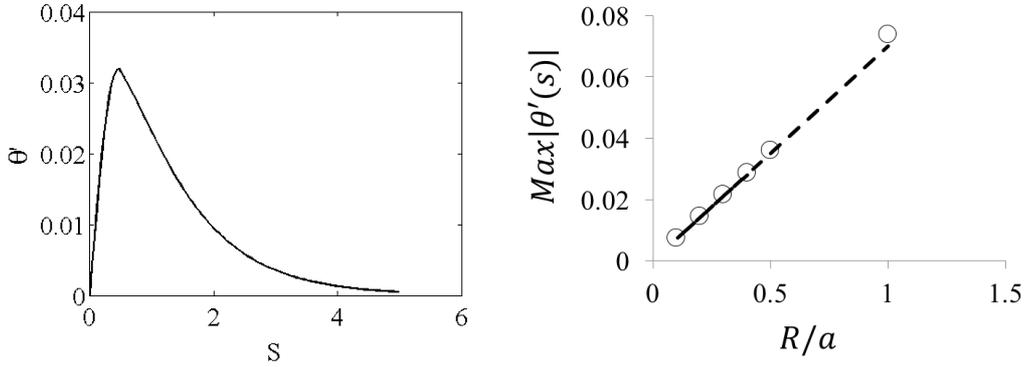

Figure 6: a) Perturbation in the local angle due to the point load for R/a=0.5. b) Maximum value of $|\theta'(S)|$ vs R/a'. The dashed line is $|\theta'(S)| = 0.07 R/a$.

The equation for the flap shape for $s \leq R/2$ is thus identical to Eq. (5), except that in this case $(a + R)$ replaces $a$. In terms of the non-dimensional curvilinear coordinate $S = s\sqrt{\mu \dot{\gamma} a / D}$, the system to be solved is thus:

$$\frac{d^2\theta}{dS^2} + \left(1 + \frac{R}{a}\right)\sin(\theta) = 0 \text{ for } S \leq \frac{1}{2} \quad (11a)$$

$$\frac{d^2\theta}{dS^2} + \sin(\theta) = 0 \text{ for } S > \frac{1}{2} \quad (11b)$$

Figure 5 illustrates a numerical solution of the system above, comparing the rotation angle $\theta(S)$ obtained for $f = 0$ with the corresponding value obtained for $f = \mu \dot{\gamma} RW$. A relatively large ratio $\frac{R}{a} = 0.7$ is chosen to make the effect of $f$ clearer on the graph. The deviation in $\theta(S)$ due to $f$ is small, and decays as $S$ increases. Because of this small deviation, it is useful to decompose $\theta$ as $\theta = \theta_0 + \theta'$ where $\theta_0$ is the unperturbed solution. Figure 6a illustrates how $\theta'$ varies with S for $\frac{R}{a} = 0.5$. The perturbation has a maximum near $S = 1$, and decays to negligible values for $S \cong 6$. Figure 6b shows how the maximum value of $\theta'$ changes with $R/a$. This figure shows that $\theta' \propto R/a$ when $R/a$ is sufficiently small. The numerical data suggests

$$\theta'(S) = \frac{kR}{a} g(S) \quad (12)$$

where $k \approx 0.07$ and $g$ is a non-dimensional function whose maximum is 1. Up to first order in $\theta'$ the bending energy can be written as

$$E \cong \frac{1}{2} D \frac{W}{R} \int_0^\infty \left(\frac{d\theta_0}{dS}\right)^2 dS + D \frac{W}{R} \int_0^\infty \left(\frac{d\theta_0}{dS}\right)\left(\frac{d\theta'}{dS}\right) dS \quad (13)$$

The first integral is the bending energy contribution appearing in Eq. 8. The second integral is, to leading order, the change in $E$ due to $f$. By observing that that the integral from 0 to ∞ converges for values of $S$ of order 1, the order of magnitude of this second term can be estimated as

$$D \frac{W}{R} \int_0^\infty \left(\frac{d\theta_0}{dS}\right)\left(\frac{d\theta'}{dS}\right) dS \sim D \frac{W}{R} k \left(\frac{R}{a}\right) \quad (14)$$

This expression shows that the bending energy contribution originating from the normal load on the fold is, to leading order, independent of $R$ and $\dot{\gamma}$.

Using Eq. (14), Eq. (7) can be written as

$$a^{-1/2}\sqrt{\mu \dot{\gamma} D} + \left(\Gamma - c_L k \frac{D}{a^2}\right) = 2\mu \dot{\gamma} a \quad (15)$$

where the constant $c_L$ is a numerical prefactor. By comparing Eqs. (15) and (7), we can see from this equation that the leading-order effect of $f$ is to reduce the critical shear rate. The net effect is analogous to reducing $\Gamma$, by an amount that depends on $D$ and $a$.

The second-order term neglected in Eq. (13), together with the corresponding external work done by $f$, would give rise to a bending energy contribution scaling as $DWk^2R/a$. This correction would translate to a term $O\left(\frac{k^2}{\sqrt{\frac{\dot{\gamma}\mu a^3}{D}}}\right)$ in Eq. 10, much smaller than the retained bending energy terms.

We have initially assumed that the normal force on the fold is $\mu \dot{\gamma} RW$. The correct prefactor, and thus $c_L$, will depend on the detailed fluid dynamics of the problem. The drag force per unit width on a cylinder of radius $R$ attached to a wall in shear flow is $4\pi \mu \dot{\gamma} R$ (Davis, 1977). If the drag on the fold is assumed to be half that on a cylinder, then $f = 2\pi \mu \dot{\gamma} RW$. Based on this estimate, the numerical prefactors should be $2\pi$ larger than assumed so far.

The effect of the normal force on the fold is typically small, because $R \ll a$. However, it could become important if the tangential force on the flap is reduced. Carbon 2D nanomaterials in contact with water exhibit relatively large slip velocities (with slip lengths in the range 10-80nm (Falk, 2010; Thomas, 2008)), meaning that the no-slip condition at the solid-liquid interface is not satisfied identically. If a large velocity slip is present, the tangential stress on the flap will be reduced from the value $\dot{\gamma}\mu$ assumed in our model. On the other hand, the value of $f$ – being related to a normal force – should not depend strongly on the slip length (a finite normal force on the fold is expected even for infinite slip lengths). In this situation, the effect of the normal force on the fold could become significant.

Modelling the force using a constant distributed force in the horizontal direction, rather than a point force, does not change the order of magnitude of the estimate presented in the current section. Indeed, in this case $\frac{d\boldsymbol{F}}{ds} = -\mu\dot{\gamma}\boldsymbol{e}_x$, which integrated gives $\boldsymbol{F} = \mu\dot{\gamma}(s^* + a - s)\boldsymbol{e}_x$, where $s^* \cong 2R$ is the location where the flap starts becoming horizontal. Inserting into the equation for the moment and normalising gives $\frac{d^2\theta}{dS^2} + \left(1 + \frac{s^*}{a} - \frac{s}{a}\right)\sin(\theta) = 0$ for $s < s^*$. The essential difference with Eq. (11a) is the term containing $\sin\theta$ multiplied by $s$; this difference makes the equation not easily solvable by direct integration (Rohde, 1953). However, because the term in parenthesis is smaller than $\left(1 + \frac{s^*}{a}\right)$, and $s^* \cong 2R$, the equation is bound from above by Eq. (11a), if we choose $f = 2\mu\dot\gamma RW$ instead of $f = \mu\dot\gamma RW$. Differences between the point force and distributed force prediction are thus within the uncertainties, which we have discussed above, in the value of the drag force $f$.

*3.3 Effect of the solvent on adhesion strength*

The parameter $\Gamma$ is the energy of adhesion per unit area of contact. In vacuum or in an inert gas, the work to separate the surfaces is twice the surface tension of the solid, $\Gamma = 2\gamma_{so}$ (Lawn, 1993).

It is well known that the presence of a liquid can reduce adhesion. Johnson, Kendall and Roberts (Johnson, 1971) carried out lubricated adhesion experiments with rubber spheres, both in air and in liquids. They found that immersion of the surfaces in water reduced the adhesion between the spheres. When the contact was immersed in a 0.01 molar solution of sodium dodecyl sulphate (SDS) the results closely agreed with the Hertz theory down to the lowest loads measured, indicating that adhesion was practically suppressed. Haidara et al. (Haidara, 1995) studied the adhesion of semi spherical PDMS lenses on flat PDMS surfaces. They also found that the presence of surfactants reduced the size of the contact region. The deformations resulting on contacting small (1-2 mm) semispherical lenses of elastomeric poly-(dimethylsiloxane)(PDMS) with the flat sheets of this material were measured in air and in mixtures of water and methanol by Chaudhury and Whitesides (Chaudhury, 1991). They found that the adhesion between PDMS surfaces was strongest in water, and decreased as the hydrophobicity of the medium decreased. Van Engers et al (van Engers, 2017) using a method conceptually similar to that of Johnson, Kendall and Roberts measured experimentally the graphene-graphene interfacial energy $\gamma_{so}$, obtaining 115 mN/m in inert gas, 83 in water, 29 in sodium cholate, a surfactant. The value obtained for graphene in an inert gas is close to the value 110mN/m previously reported for graphene-graphite interaction by Wang et al. (Wang, 2016). Shih et al (Shih, 2010) carried out molecular dynamics of the interaction between rigid graphene sheets in a variety of solvents, and obtained a minimum in the interaction energy of $250 kJ mol^{-1} nm^{-2}$ for water and around $100 kJ mol^{-1} nm^{-2}$ for N-methylpyrrolidone (NMP). Because the reduction in cohesive stresses can be related to a reduction in the depth of the potential energy well for the molecular bonds near the crack tip (Stoloff, 1963; Lawn, 1993), this last data suggests that the value of the surface tension for water and NMP are of comparable order of magnitude (despite the fact that NMP is considered a much better solvent for graphene than water!).

From the data above there is evidence to suggest that: i) the intrinsic value of $\Gamma$ corresponding to vacuum or an inert gas is in the range 0.20-0.25N/m; ii) "good" solvents can reduce $\Gamma$, but probably not by several orders of magnitude. To compare with experimental data we will assume that $\Gamma$ in typical liquid solvents ranges from $0.01N/m$ to $0.1N/m$, with the lower value being characteristic of a good solvent.

*3.4 Critical shear rate: comparison with experimental data*

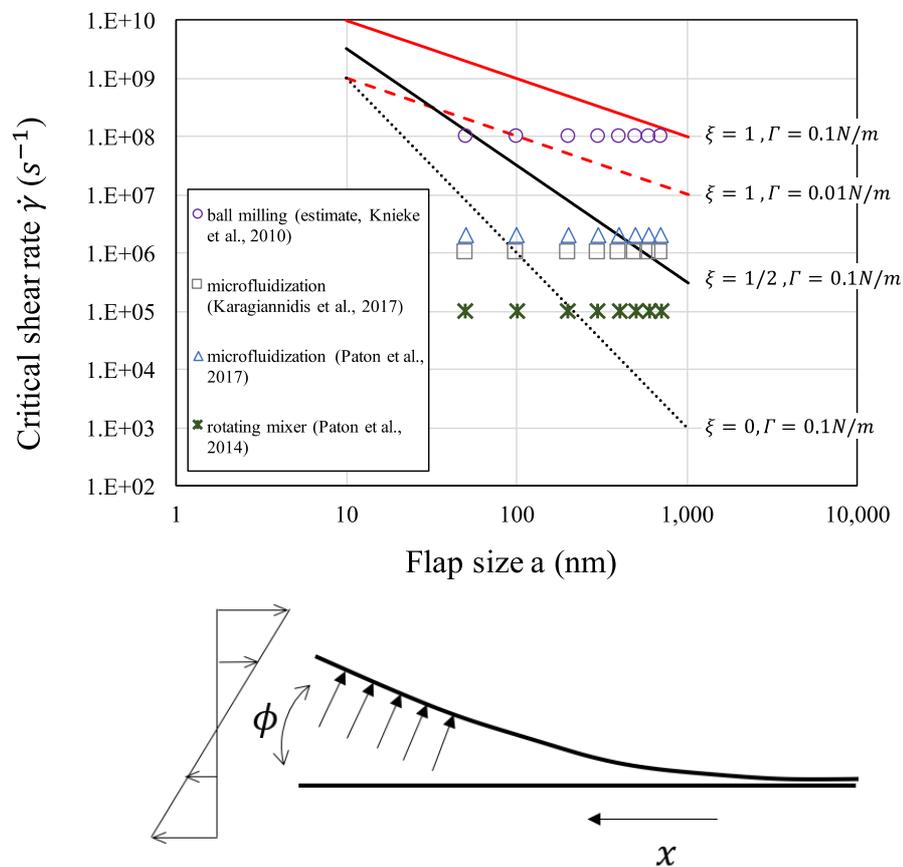

*Figure 7: a) Critical shear rate for exfoliation as a function of flaw size (in nanometers) for different scaling exponents, assuming $D = 10^{-18} J$. Experimental references: Knieke et al. (Knieke, 2010); Karagiannidis et al. (Karagiannidis, 2017); Paton et al. (Paton, 2014) ; Paton et al., 2017 (Paton, 2017). b) In the "cleavage" configuration the fluid stresses act normal to the flap, and the opening angle $\phi$ is small.*

We have seen that if the microscopic peeling configuration is as in Fig. 2 the order of magnitude of the critical shear rate is given by $\dot\gamma \approx \frac{\Gamma}{2\mu a}$, i.e. Eq. 3 with $\xi = 1$. Figure 7a compares $\dot\gamma = \frac{\Gamma}{2\mu a}$ curves against experimental data for $\mu = 10^{-3} Pa \cdot s$ and two surface energy values: $\Gamma = 0.1 N/m$ and $\Gamma = 0.01 N/m$. Each experimental case is denoted by a horizontal row of symbols corresponding to several values of $a$. This was done because the value of $a$ in experiments is unknown. Experimental cases correspond to turbulent exfoliation in a rotor-stator mixer

(Paton, 2014), turbulent exfoliation in microfluidization (Paton, 2017; Karagiannidis, 2017) and wet ball milling (Knieke, 2010). In the case of wet ball milling, a typical stress energy SE~0.134$\mu J$ was reported for ball diameter $d_{gm} = 100\ \mu m$ and rotation rate 1500rpm. The critical stress energy can be converted to a stress $\tau$ of the order of $10^5 Pa$ assuming that SE is dissipated within a contact region of volume $\sim d_{gm}^3$. With $\tau = \mu\dot\gamma$ and viscosity $\mu = 10^{-3} Pa \cdot s$, a stress of $10^5 Pa$ corresponds to an equivalent critical shear rate value of $10^8 s^{-1}$. This is the estimated value reported in Fig. 7a.

The graph that for realistic values of Γ and for a range of realistic values of flaw size (we can estimate that for particles of length $L = 1\mu m$ the flaw size ranges roughly from $50 nm$ to $500 nm$) the expression $\dot\gamma \approx \frac{\Gamma}{2\mu a}$ largely overestimates the experimental data for turbulent exfoliation (i.e. exfoliation in rotor-stator mixers or microfluidisation). Even for Γ=0.01N/m, which is smaller than any measured value for the adhesion energy of graphene (see discussion in Sec. 3.3), the critical shear rate is much larger than what reported for rotating mixer and microfluidisation. Instead, the values predicted by $\dot\gamma \approx \frac{\Gamma}{2\mu a}$ are reasonably close to those estimated for ball milling, a technique that allows large stresses to be produced.

*3.5 An alternative model: flow-induced cleavage (for small opening angles)*

From Fig. 7a it can be seen that if an exponent $\xi = 1/2$ is assumed, a much closer agreement with the experimental data for rotating mixer and microfluidisation approaches is achieved. An exponent of $1/2$ can be obtained if we assume that the bending of the flap is caused by a *normal* stress on the flap of order of $\mu\dot\gamma$, as in the "cleavage" configuration illustrated in Fig. 7b. In this case the flap would be subject to a force $\sim \mu\dot\gamma a W$ acting on a lever arm $\sim a$. Equating the corresponding external moment $\sim \mu\dot\gamma a^2 W$ to the bending moment $DW\sqrt{\Gamma/D}$ required for fracture initiation (Obreimoff, 1930) leads to $\frac{\mu\dot\gamma a^3}{D} \sim (\Gamma a^2/D)^{1/2}$. The same result can be obtained by considering the analytical solution for the displacement of a cantilever beam of length $a$ subject to a constant normal load $\mu\dot\gamma$ (Timošenko, 1940):

$$w = \frac{\mu\dot\gamma}{24D}(x-a)^2(x^2 + 2ax + 3a^2) \quad (16)$$

The bending energy per unit width $U = \frac{D}{2}\int \left(\frac{d^2 w}{dx^2}\right)^2 dx$ is quadratic in $\mu\dot\gamma$. The strain energy release rate $G = \frac{dU}{da}$, with units of energy per unit area, is proportional to $\frac{(\mu\dot\gamma)^2 a^4}{D}$. Equating the strain energy release rate to Γ (Griffith's criterion (Lawn, 1993)) gives again the scaling

$$\frac{\mu\dot\gamma a^3}{D} \sim (\Gamma a^2/D)^{1/2} \quad (17)$$

An intuitive explanation for why the critical shear rate in the "cleavage" configuration is smaller than in the "$\pi$ - peel" configuration is that even if the forces applied to the flap are the same in both cases, the lever arm in the "$\pi$ - peel" configuration is of $O\left(\frac{R}{a}\right)$ smaller than in the "cleavage" configuration. For a given value of $a$, reaching the critical bending moment in the "cleavage" configuration therefore requires as smaller value of $\dot\gamma$.

Equation (17) is valid provided that the displacement of the flap is small. For larger displacements, one needs to account for two factors, which are extensively discussed in a recent paper by the author (Salussolia, et al., 2019) : i) even in the case of uniform pressure applied to the flap, the direction of load depends on the normal to the flap, which itself depends on the shape of the flap (i.e. the load is follower); ii) as the flap displacement increases, the pressure on the flap increases, essentially because more area of the flap is exposed to flow. Factor ii) makes the dependence on $\Gamma a^2/D$ weaker, reducing effectively the power-law exponent significantly below 1/2. The inclusion of large-deformation effects does not change this conclusion, although it changes somewhat the value of the critical shear rate (without a change in order of magnitude of this quantity).

*3.5 Time scales of microstructural rearrangement: towards exfoliation kinetics*

Once the critical shear rate is reached, the fracture will propagate and exfoliation will occur. But exfoliation is not an instantaneous event. If this was the case, at a critical shear rate one would obtain complete exfoliation of all the microplates. This is not observed in practice. For instance, Paton and co-workers found that the concentration of exfoliated material follows a rather slow kinetics, apparently governed by a power-law of time (Paton, 2014). While explaining the emergence of this power-law requires tools that go beyond simple micromechanics, it is instructive to consider examples of dissipation processes occurring in the vicinity of the crack tip that could make the exfoliation mechanics depend on time.

An important dissipative process is due to the viscosity of the solvent. As the crack of the interlayer interface propagates, fluid must be drawn in from the surroundings towards the crack tip. Because the gap distance between the layers is small in the crack region, the viscous dissipation can be substantial (Rieutord, 2005) (Lister, 2013). The rate of dissipation per unit volume of fluid is

$$\varepsilon \sim \mu \left(\frac{U(x)}{h(x)}\right)^2 \quad (17)$$

where $h(x)$ is the gap height at a position $x$, and $U(x)$ is the characteristic fluid velocity at $x$ (Eggers, 2015). The total power per unit width dissipated in the fluid can be estimated to be of the order of (Rieutord, 2005)

$$\frac{P_d}{W} = \int_{x_{min}}^{x_{max}} \int_0^h \varepsilon \, dxdy \sim \mu \frac{V^2}{H^2}(H\ell_D) \quad (18)$$

where $V$ is the crack tip speed. Here $H$ and $\ell_D$ are the characteristic height and length of the wedge-like region in the vicinity of the crack tip where the bulk of the viscous dissipation occurs. The high-dissipation region is limited by the coordinates $x_{min}$ and $x_{max}$. If there is no external force applied to the flap, the power to drive the crack motion against the viscous dissipation is provided by the adhesion force, $\frac{P_d}{W} = \Gamma V$, leading to (Rieutord, 2005)

$$V \sim \frac{\Gamma}{\mu} \frac{H}{\ell_D} \quad (19)$$

In the "$\pi$ - peel" case examined in the current paper, an external force per unit width $\mu\dot{\gamma}a$ is applied to the flap and the corresponding driving power is $\frac{P_d}{W} = (2\mu\dot{\gamma}a - \Gamma)V$. Hence, the crack velocity, for the case in which viscous dissipation is the only effect resisting the motion of the crack, can be estimated to be of the order of

$$V_{viscous} \sim \frac{(2\mu\dot{\gamma}a - \Gamma)}{\mu} \frac{H}{\ell_D} \quad (20)$$

The inverse of the geometrical ratio $\frac{H}{\ell_D}$ is equal to the integral of $h(x)$ from $x_{min}$ to $x_{max}$ (Rieutord, 2005). Because the region near the crack tip is thin and slender, we have $\frac{H}{\ell_D} \ll 1$. Thus, $V_{viscous} \ll \frac{(2\mu\dot{\gamma}a - \Gamma)}{\mu}$.

A second important dissipative process is caused by the time-dependent rupture of the molecular bonds in the adhesion zone. A model for such process assumes that the rupture rate is governed by Maxwell-Boltzmann statistics. For small deviations from Griffith's condition, this model gives an exponential dependence of the crack velocity on the difference between the strain energy release rate $G$ and the energy of adhesion per unit area of contact $2\gamma_{so}$ (Lawn, 1993):

$$V = 2v_0 a_0 \exp\left(-\frac{\Delta U°}{kT}\right) \sinh\left(\alpha \frac{G - 2\gamma_{so}}{kT}\right) \quad (21)$$

Here, $a_0$ is the characteristic lattice dimension, $v_0$ is a molecular frequency (typically a few THz), $\alpha$ is an activation area, $kT$ is the thermal energy, $\gamma_{so}$ is the surface tension of the solid (Sec. 3.3) and $\Delta U°$ is a quiescent activation energy. In the "$\pi$ - peel" configuration, the external work $2\mu\dot{\gamma}a$ takes the place of $G$. Using the definition $\Gamma = 2\gamma_{so}$ the following expression crack velocity expression is obtained:

$$V_{br} \sim V_{br,0} \sinh\left(\alpha \frac{2\mu\dot{\gamma}a - \Gamma}{kT}\right) \quad (22)$$

where

$$V_{br,0} = 2v_0 a_0 \exp\left(-\frac{\Delta U°}{kT}\right) \quad (23)$$

The molecular velocity $v_0 a_0$ is of the order of 100m/s-1km/s, i.e., sonic velocities. The energy barrier $\Delta U°$ is typically taken to be of the order of $10kT$ (Brochard-Wyart, 2003). Setting $\Delta U° = 25kT$, gives values of $V_{br,0}$ in the range $1 - 10 nm/s$. The effect of the solvent (through chemically-assisted bond rupture) can be accounted for through the dependence of $\Gamma$ on the solvent, and by appropriately modifying the activation terms (Lawn, 1993) (Stoloff, 1963).

We can use equations (20) and (22) to make some considerations regarding the relative importance of viscous dissipation and dissipation due to bond rupture. Let us assume that we are carrying out exfoliation using a shear rate of the order of the critical one, a situation that is expected to hold in practice. Because $\frac{2\mu\dot{\gamma}a}{\Gamma} = O(1)$ and $\frac{H}{\ell_D} < 1$, an upper bound for $V_{viscous}$ is the "capillary velocity" $\frac{\Gamma}{\mu}$. This quantity is approximately 100 m/s for solvents having the viscosity of water (using a reference value $\Gamma$=0.1N/m). The velocity due to (interlayer) molecular bond rupture can be estimated to be, typically, larger than this value. This is because for $a_0$ in the nanoscopic range the energy scale $\Gamma\alpha$ is typically larger than the energy barrier $\Delta U°$ (taking $\alpha = 1nm^2$ and $\Gamma$=0.1N/m, we get $\Gamma\alpha \approx 250kT$). As a result, the growing hyperbolic sine term is much larger than the decaying term $\exp\left(-\frac{\Delta U°}{kT}\right)$. Noting that the "capillary velocity" $\frac{\Gamma}{\mu}$ is roughly comparable in magnitude to $v_0 a_0$, we have $V_{viscous} \ll V_{br}$.

Thus, viscous dissipation can be very important. The relative importance with respect to dissipation due to bond rupture (which we estimate to be subdominant) will depend on the very specific values of $\alpha$ and $\Delta U°$. Recent measurements of self-tearing and peeling of graphene sheets from solid substrates suggest a method to calculate these parameters (Annett, 2016).

In the regime of viscous dominated dissipation, the characteristic time required to complete peeling of a flap from a particle of lateral size $L$ is

$$T_v = \frac{L}{V_{viscous}} \sim \frac{\mu}{(2\mu\dot{\gamma}L - \Gamma)} \frac{\ell_D}{H} \quad (24)$$

For values of the applied shear rate much larger than the critical one the viscous peeling time scales proportionally to the inverse shear rate, $T_v \sim \dot{\gamma}^{-1} \frac{\ell_D}{2H}$. This time scale depends on the viscosity only through the dependence of the ratio $\frac{\ell_D}{2H}$ on hydrodynamics. Because $\frac{\ell_D}{H} \gg 1$ we expect $T_v \gg \dot{\gamma}^{-1}$. The viscous peeling time scale is proportional to the local shear rate, with a large numerical prefactor.

In this section we have focused on the "$\pi$ - peel" configuration. The arguments proposed can be extended to the "cleavage' configuration by replacing the work of the external force per unit width with the strain energy release rate.

## 4. *Conclusions: frontiers in the hydrodynamics of 2D nanomaterials*

We have analysed theoretically a simple model of hydrodynamic peeling, by focusing mostly on the "$\pi$ - peel" configuration (Fig. 2). Based on our analysis, the shear rates that one needs to apply to the fluid for this configuration to result in exfoliation are rather large, of the order of $10^8 - 10^9 s^{-1}$. These large shear rates are available in exfoliation methods that produce large stresses on the particle, such as ball milling, but not in most published exfoliation approaches in which the shear rate is produced by conventional turbulence (as in rotor-stator shear mixing or microfluidisation approaches). The fact that in rotor-stator or microfluidisation approaches exfoliation is observed for shear rates in the range $10^4 - 10^6 s^{-1}$ suggests that alternative microscale exfoliation configurations are likely to dominate the exfoliation micromechanics in these approaches, at least in the initial stages of exfoliation.

We suggest that, due to its stronger dependence on the flaw size $a$, a "cleavage" configuration (Fig. 7b) can yield critical shear rate values much smaller than those predicted for "$\pi$ - peel", closer to the values observed experimentally. While for "$\pi$ - peel" the critical shear rate decays as $\frac{1}{a}$ (Eq. 9), for cleavage under a constant pressure $\mu\dot{\gamma}$ the critical shear rate decays as $\frac{1}{a^2}$ (Eq. 17). We anticipate that if the pressure is not constant, but depends on the configuration of the wedge creating the fracture, the dependence on $a$ can be even stronger. This effect results in even lower values of the critical shear rate. The analysis of this case is the subject of a separate paper (Salussolia, et al., 2019).

This initial study on the hydrodynamics of 2D materials exfoliation only begins to uncover the complexity of the exfoliation process as seen at the scale of each particle. To better understand the micromechanics of liquid-phase exfoliation, future computational investigations must consider the fully coupled fluid-structure interaction problem and include atomistic details. In the wedge region, near the crack tip, the liquid is strongly confined, with gaps of nanometric dimensions. Hence the problem severely challenges the use of continuum approaches. For carbon nanomaterials in contact with water, hydrodynamic slip characterises the flow behavior at the solid-liquid surface (Tocci, 2014) (Striolo, 2016), with slip lengths of the order of a few tens of nanometers (Falk, 2010; Thomas, 2008). Our work provides estimates for geometric quantities – such as the shear rate dependent radius of curvature of the fold $R$ - that are obtained neglecting molecular effects. These quantities can be used for reference to evaluate the limitations of continuum treatments. Molecular dynamics studies for configurations indicated in the current paper could provide insights into the general process of liquid intercalation under flow, a topic that could lead to the design of improved solvents for the exfoliation of 2D nanomaterials.

Another important topic is the prediction of the kinetics of exfoliation. We have estimated that the time-scale for microscopic peeling can in many cases be controlled by viscous dissipation. For "$\pi$ - peel" at shear rates largely exceeding the critical one, the time-scale for viscous peeling has been found to scale proportionally to $\dot{\gamma}^{-1}$ (with a large prefactor, $\frac{\ell_D}{H} \gg 1$, see Eq. 24). This time scale has to be compared to two dynamic time-scales: the time-scale of particle rotation (of the order of $\dot{\gamma}^{-1}\Lambda$, where $\Lambda$ is the aspect ratio (MLA Challabotla, 2015)), and the time scale of permanence in a turbulent structure (which can be estimated to be of the order of $\dot{\gamma}^{-1}$ (Babler, 2012). The inter-play of these different time scales should give rise to a rich phase diagram. Uncovering the features of this diagram, through a comparison of multi-particle simulations and experiments (Voth, 2015), should be of interest for researchers investigating the statistical physics of complex systems, and is a necessary step towards predicting the yield of industrial-scale exfoliation processes.

Theoretical modelling of exfoliation processes in sheared liquids is an important and currently largely unexplored frontier in carbon nanomaterials research. Understanding how graphene "breaks" under the action of fluid dynamic forces has implications not only for large-scale graphene production but also for quantifying the time-evolving size and thickness distribution of graphene (or any other 2D material) *during* its transport. In industrial liquids such as lubricants and paints, the performance of a graphene additive whose size depends on time - because of fragmentation or exfoliation - will also depend on time. Size and shape are dominant variables affecting how a nanoparticle interacts with a biological cell, and the prediction of these variables during flow is thus essential to evaluate toxicological effects on human health or the environment (Wick, 2014). Our work on the modelling of graphene hydrodynamics, sponsored by the European Research Council, lays basic theoretical building blocks that are necessary to develop the next generation of predictive multiscale software for fragmentation, exfoliation and liquid processing of 2D nanomaterials.

*Acknowledgements*

Funding from ERC Starting Grant FLEXNANOFLOW (n. 715475) is gratefully acknowledged. The author would like to thank Dr. E. Barbieri, and G. Salussolia for useful scientific discussions.